\documentstyle[12pt,psfig]{article}

\begin{document}

\newcommand{\alt}{\mathbin{\lower 3pt\hbox
   {$\rlap{\raise 5pt\hbox{$\char'074$}}\mathchar"7218$}}}
\newcommand{\agt}{\mathbin{\lower 3pt\hbox
   {$\rlap{\raise 5pt\hbox{$\char'076$}}\mathchar"7218$}}}

\begin{center}{\Large \bf
Upper critical dimension in the scaling theory of localization} \\
\vskip.25in
{I.~M.~Suslov}

{\it
P.~L.~Kapitza Institute for Physical Problems, Russian Academy of
Sciences,\\ 117334, Moscow, Russia
} %
\end{center}

\begin{abstract}                % DON'T CHANGE THIS LINE
It is argued that the Thouless number $g(L)$ is not the only parameter
relevant in scale transformations, and that the second parameter connected
with off-diagonal disorder should be introduced. A two-parameter scaling
theory is suggested that explains a phenomenon of the upper critical
dimension  from the viewpoint of scaling ideas.

\end{abstract}

\vspace{5mm}
PACS numbers 03.65.-w, 11.10.Hi, 71.23.An
\vspace{5mm}

The one-parameter scaling hypothesis \cite{1} provides the basis for the
contemporary theory of localization. Its justification is still actual
 \cite{2} and may require  more accurate definitions of the scaling
variables as well as lead to a restriction of the range of applicability.
Here we discuss modifications of the scaling hypothesis which are inevitable
in high dimensions.

The scaling consideration is usually applied only for space dimensions $d$
being in the interval between $d_{c1}$ and $d_{c2}$, the lower and the upper
critical dimensions \cite{14}. There are no doubts that $d_{c1}=2$ \,
\cite{1}, whereas the value of $d_{c2}$ was disputable for many years
\cite{3,4,5}.  Recently it was established in the
author's series of papers \cite{6} that $d_{c2}=4$ for the problem of the
density of states (defined by the average Green function $\langle G
\rangle$).  The value $d_{c2}=4$ is distinguished from the viewpoint of
renormalizability:  the theory is nonrenormalizable for dimensions $d>4$ and
the cutoff is necessary on the atomic scale for high momenta. Since the
atomic scale cannot be left out of consideration, no scale-invariance is
possible.  The same argument can also be applied to the problem of
conductivity defined by the correlator $\langle G^R G^A \rangle$. This is
confirmed by author's "symmetry theory" \cite{7}, that reproduces on a
rigorous level the results of Vollhardt and W$\ddot o$lfle \cite{8} and gives
the values of critical exponents (claimed to be exact in Ref.~5),  which are
consistent with the one-parameter scaling theory only for $d<4$.

On the other hand, the recent assertions that $d_{c2}=\infty$ \,\cite{5}
have some grounds: {\it there is no place for the upper critical dimension
in the one-parameter scaling theory}. Overall, there are certain drawbacks
in the existing physical picture of localization.  Another question to be
cleared out is related to the mechanism responsible for violation of the
Wegner relation $s=(d-2)\nu$ for $d>4$ \, \cite{7,8,4}.  The present paper is
aimed to fill these gaps in our knowledge.

The scaling theory \cite{1} is based on Thouless's scaling considerations
\cite{10}. A  disordered system, that is described by the
Anderson model with the overlap integrals $J$ between the nearest neighbours
and with the spread $W$ of cite levels, is divided into blocks of size $L$.
In the absence of interaction between the blocks, each of them has a random
system of energy levels with a characteristic spacing $ \Delta (L) \sim J
\left ( a_0/L \right)^d $, where $a_0$ is the lattice constant. When
the interaction between the blocks is "switched on",
the states of neighbouring blocks become coupled and the
corresponding matrix elements become nonzero. This gybridization
is the most essential for the states nearest in energy, and should be taken
into account first of all. If the level nearest
to the considered energy $E$ is selected in each block, we obtain an
effective Anderson model with the spread of levels $W(L) \sim \Delta (L)$ and
overlap integrals $J(L)$ determined by corresponding matrix elements.  This
model describes the system on the scale larger than $L$, and is characterized
by the Thouless parameter
$$
g(L)=\frac{J(L)}{W(L)} \quad, \eqno(1)
$$
which
is also equal to the dimensionless conductance of the block \cite{1}.
Repetition of the same consideration for the effective Anderson model
constitutes the principal algorithm for evaluation of $g(bL)$ with integer
$b$ when value $g(L)$ is known: $ g(bL)=F \left( b, g(L) \right)$.  Taking
the limit $b\to 1$  in this relation, results in the Gell-Mann and Low
equation \cite{1}
$$
\frac{d \ln g}{d \ln L} =\beta (g) \quad.  \eqno(2)
$$
For $d>2$, there exists a phase transition point $g_c$, defined by the
condition $\beta(g_c)=0$, and, in a vicinity of the transition, the
conductivity $\sigma$ and localization length $\xi$ have the following
behavior
$$
\sigma\sim (g_0-g_c)^{s} \quad, \qquad \xi \sim (g_c-g_0)^{-\nu} \quad.
 \eqno(3)
$$
Here $g_0$ is the value of $g(L)$ on the scale $L\sim a_0$, and the critical
exponents  are given by $1/\nu = g_c \beta'(g_c) \quad$  and $s=(d-2)\nu$. \,
\cite{1}

The above consideration relies heavily on the assumption  that $g(L)$ is the
only relevant parameter in the scale transformations. We shall show that,
in general, it is not the case.  In order to see it, let us assume that the
typical wave function of localized states has the form
$$
|\Psi(r)|
\sim \left \{ \begin{array}{cc} r^{-\zeta}\,\, , & r \ll \xi \\ \exp(-r/\xi)
\,\, ,  & r \gg \xi \end{array} \right . \quad ,
\eqno(4)
$$
where the
 exponent $\zeta$ goes to infinity for large $d$. There are some reasons for
such an assumption: (i) the optimal fluctuation method \cite{11} results in
Eq.~(4) with $\zeta=d-2$ in the range of deep localization, and (ii) an
analogous behavior for the critical region can be guessed from the
$d$-dependence of the exponent $\eta$ of density correlator \cite{12}.
A large value of $\zeta$ means that the eigenfunctions of separate blocks in
Thouless's construction are well localized on the scale $L\ll \xi$ (Fig.~1).
Consequently, strong off-diagonal disorder appears: f.e. the overlap
integral between the states 1 and 2 is much smaller than one between  states
3 and 4.  With the increase of $\zeta$ we approach the well-known situation
of topological disorder in the system of impurities with exponential overlap
(Ref.~13). So a catastrophe, viz.  localization due to the pure
off-diagonal disorder, becomes possible.  It can even occur for $W(L)=0$,
when the Thouless parameter is infinite and cannot play any role. Therefore,
it is reasonable to suggest that gybridization of the block states is
determined by some other parameter connected with the off-diagonal disorder.

Let us suppose that a disordered system is characterized by two parameters
$$
g(L)=\frac{J(L)}{W(L)}\quad, \qquad
\varphi(L)=\frac{\delta J(L)}{J(L)}\quad, \eqno(5)
$$
where $\delta J(L)$ is the fluctuation of overlap integrals. The
boundary $AB$ (Fig.~2) between localized and extended states should then be
situated at $g\sim 1$  for $\varphi=0$ and go to infinity at some
critical point $\varphi_c$,
in accordance with the possibility of
localization due to the pure off-diagonal disorder.
%thereby pointing out the possibility of localization for the pure
%off-diagonal disorder.
In the course of scale transformations, one point
of the $(g, \varphi)$-plane turns into another point of this plane, and one
point of line $AB$ turns  into another point of this line.

To return to the one-parameter scaling, it is sufficient to assume  the
existence of a fixed point $F$, that is stable on the critical surface ($AB$)
and unstable beyond it (Ref.~3, Ch.~6).  The point $F$ is of saddle-type
and characterized by two asymptotes, $AB$ and $CD$ (Fig.~2,$a$).  The
movement in the $(g, \varphi)$-plane can be roughly divided into two stages:
relaxation to line $CD$ on some scale $L_0$ and evolution along $CD$ on the
scale $\xi$, which is arbitrarily large near the phase transition. For $L\gg
L_0$, the whole $(g, \varphi)$-plane reduces to line $CD$, and the position
on the latter is uniquely determined by the Thouless parameter $g(L)$.
Thereby we return to the usual picture of localization, and we assume it to
be valid for low dimensions.

Let us now suppose that  for high dimensions  there is no fixed point on the
critical line $AB$ (Fig.~2,$b$).  If a system is in a critical point, then it
moves upwards along this line (the movement downwards means that the
off-diagonal disorder dissapears asymptotically, and contradicts
to the previous arguments). Consequently, this implies that in the critical
point parameter $g(L)$ increases (in contrast to $g(L)=const$ in the previous
scenario): it does not mean that degree of gybridization  grows but indicates
that the diagonal disorder transforms to off-diagonal one. In the
metallic phase, $g(L)$ is represented by a more rapid dependence $\sim
\sigma L^{d-2}$ (Ref.~1), and, in the localized phase, it exhibits
non-monotonic behavior, i.e. increasing for $L\alt\xi$ and decreasing for
$L\agt\xi$.

The first scenario is changed by the second one at  some critical value of
$d$, which we identify with $d_{c2}$. To obtain the phemomenological
description of such a bifurcation we introduce a new variable
$h=F(g,\varphi)$, so that in the $(g,h)$-plane the line  $AB$ has
behavior $g\sim h$ for large $g,h$ (the critical line will then have
regular projections on both axes), and the other asymptote $CD$ becomes
vertical (this would simplify equations).  In the case of the two relevant
parameters, $g(L)$ and $h(L)$,
%we have in analogy with (2)
the following relations can be written down by following the usual line of
reasoning (cf. Eq.~(2)):
$$ \frac{d \ln g}{d
\ln L} =\beta (g,h)   \quad, \qquad \frac{d \ln h}{d \ln L} =\gamma (g,h)
\quad.\eqno(6)
$$
In the region of large $g$ and $h$, where the fixed point $F$ is
situated for $d$ close to $d_{c2}$, Eqs.~(6) take the form
$$ \frac{d
\ln g}{d \ln L} = (d-2)+\frac{Ah}{g}+\frac{Bh^2}{g^2}
+\frac{Ch^3}{g^3}+\ldots \equiv (d-2)+ \tilde\beta \left( \frac{g}{h}\right)
\quad,\eqno(7  )
$$
$$
\frac{d \ln h}{d \ln L} =\mu+\frac{b}{h}
\quad,\qquad\qquad\qquad\qquad \qquad\qquad\qquad\qquad\qquad\qquad\qquad
\eqno(7 b)
$$
where $\mu$ changes the sign at the point $d= d_{c2}$,
$$
\mu=\alpha (d-d_{c2}) \quad, \qquad d\to d_{c2} \eqno(8)
$$
and the following unequalities are satisfied: $\alpha>0\,,\,\,b>0\,,\,
$and$ \,
A<0\,$.  Indeed, at constant $h$ the function $\beta(g,h)$ should have all
the properties discussed in Ref.~1, and it should then be expanded
accordingly:  $\beta(g,h)= (d-2)+A_1(h)/g+A_2(h)/g^2 +\ldots \,$. In
addition, the coefficients $A_n(h)$ should have expansion in $1/h$ beginning
with $h^n$ in order to yield a root $g_c \sim h$. Keeping the leading terms
with respect to $h$ results in Eq.~(7$a$).  For $d>d_{c2}$, function
$\gamma(g,h)$ should provide the indefinite growth of $h$, which, however,
should not be faster than that of $g$. This gives unequility $0<
\gamma(g,h)< d-2\,$, suggesting that the expansion of $\gamma(g,h)$ in
$1/g,\, 1/h$ begins with zero-order:  $\gamma(g,h)=\mu+a/g+b/h+\ldots
\,$. In the case of the vertical asymptote $CD$, the fixed point $h_c$ is
independent of $g$, and thereby $a=0$. The fixed point should be stable for
$d<d_{c2}$ and  absent for $d>d_{c2}$.  This requires that $b>0$ and $\mu$ to
change the sign in the point $d_{c2}$.

The system of Eqs. (7) can be easily investigated. For  $d<d_{c2}$,
Eq.~7$b$ has a fixed point $h_c=b/|\mu|$ and replacement $g\to g
h_c$ in Eq.~(7 ) results in the one-parameter scaling description
with critical exponents given by
$$
1/\nu=g_c \tilde \beta'(g_c)\quad, \qquad
s=\nu(d-2)\quad, \qquad (d-2)+\tilde \beta(g_c)=0\quad.
\eqno(9)
$$
For $d>d_{c2}$, we have $h(L)\sim L^{\mu}$ at large  $h$, and the replacement
$g\to g L^\mu$ in Eq.~7$b$ gives
$$
1/\nu=g_c \tilde \beta'(g_c)\quad, \qquad
\eqno(10 a)
$$
$$
s=\nu(d-2-\mu)\quad, \qquad  \eqno(10 b)
$$
$$
(d-2-\mu)+\tilde \beta(g_c)=0\quad. \eqno(10 c)
$$
For $L\alt \xi$,  the Thouless parameter can be written down as follows:
$$
g(L)=g_c+(g_0-g_c)\left(L/a_0\right)^{1/\nu} \quad, \qquad
d<d_{c2} \quad, \eqno(11 a)
$$
$$
g(L)=g_c\left(L/a_0\right)^{\mu}+
     (g_0-g_c)\left(L/a_0\right)^{\mu+1/\nu} \quad, \qquad
d>d_{c2} \quad. \eqno(11 b)
$$
In the critical point $g(L)$ grows as $L^\mu \,$ for $d>d_{c2}$, thereby
leading to the violation of the Wegner relation (see Eq.~(10$\,b$)). In
general, critical exponents as functions of $d$ have cusps at $d=d_{c2}$.
Usually the critical exponents are independent of $d$ for $d>d_{c2}$.
According to Eq.~(10$\,b$) this would become possible for
$\mu =d+{\sl const}$, which results together with Eq.~(8) in
$$
\mu=d-d_{c2} \quad.  \eqno(12)
$$

The results obtained can be compared with the symmetry theory \cite{7} that
gives the same critical exponents as in Ref.~9:
$$ \nu=1/(d-2)\quad, \qquad s=1 \qquad
{\rm for} \qquad 2<d<4\quad,
$$
$$
\nu=1/2\quad, \qquad s=1 \qquad {\rm for}
\qquad d>4\quad. \eqno(13)
$$
The Wegner relation $s=\nu(d-2)$ is valid only for $d<4$ implying that
$d_{c2}=4$.  To obtain the result analogous to Eqs.~(11), we find from
Ref.~8 the diffusion constant  $D_L$ of a finite block of size $L$. It is
determined by the diffusion coefficient $D(\omega, q)$ of the infinite
 system: $\,\,D_L \sim
D\left(iD_L/L^2, L^{-1} \right) \,.$
\footnote{This relation  is not valid in the localized phase for
 $L\agt \xi$ due to the nonlocal response \cite{8}.}  Using Eqs. (112) and
(116$a$) of Ref.~8 and $g(L)\propto D_L L^{d-2}$ one can finally obtain the
folowing reation:
$$
g(L)=g_c\left(L/a_0\right)^{d-2-1/\nu}+
(g_0-g_c)\left(L/a_0\right)^{d-2} \quad. \qquad \eqno(15)
$$
This result
coincides with Eq.~(11) if Eqs.~(12) and (13) are taken into account.  Such
an agreement is nontrivial because the symmetry theory \cite{7} is based on
completely different principles without any reference to the scaling ideas.

This paper was stimulated by discussions with V.~E.~Kravtsov, A.~D.~Mirlin
and M.~V.~Feigel'man.  It is supported by  INTAS (grant 96--0580) and the
Russian Fund of Fundamental Research (grant 96-02-19527).

\newpage
\begin{figure}[p]
 \label{fig1}
 \psfig{file=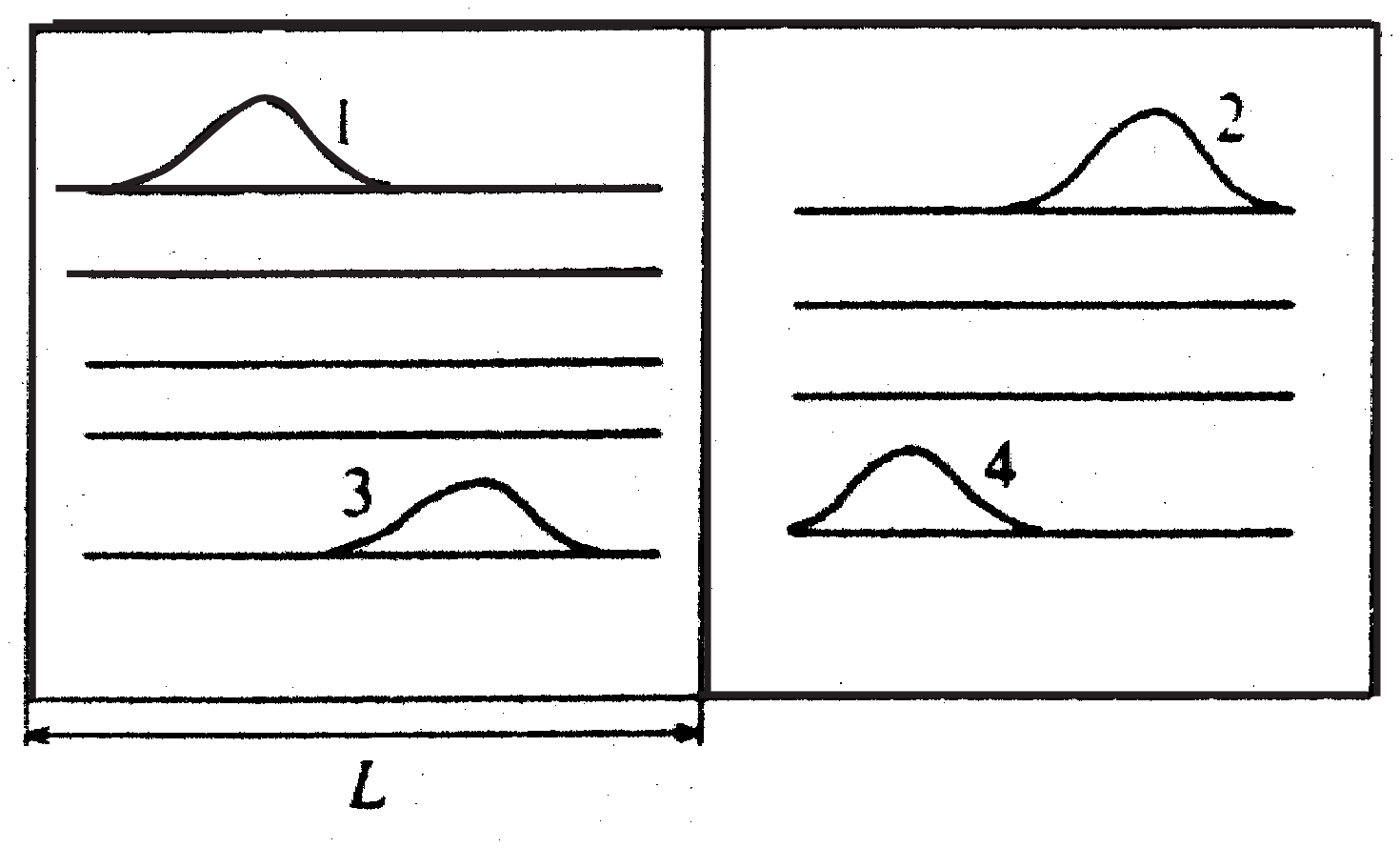,width=5.1 in}
\caption{}
\end{figure}

\begin{figure}[p]
 \label{fig2}
\psfig{file=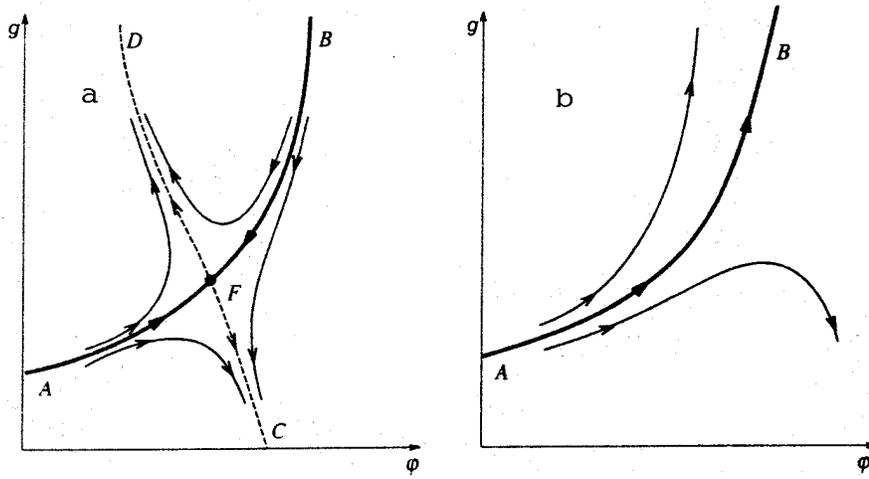,width=5.1 in}
 \caption{Flow diagram in the $(g, \varphi)$ plane: ($ $) in the case of
existence of the fixed point $F$ on the critical surface $AB$, ($b$) in the
case of its absence.
 }
\end{figure}

\end{document}